\documentclass{xray_symp}
\usepackage{graphics,psfig}

\def\erg{~{\rm erg~cm}^{-2}~{\rm s}^{-1}}
\def\ergs{~{\rm erg~s}^{-1}}

\begin{document}
\title{The ROSAT Deep Surveys}
\author{G. Hasinger\inst{1} \and I. Lehmann\inst{1} \and R. Giacconi\inst{2}
       \and M. Schmidt\inst{3} \and J. Tr\"umper\inst{4} and 
       G. Zamorani\inst{5}}
\institute{Astrophysikalisches Institut Potsdam,
       An der Sternwarte 16, 14482 Potsdam, Germany
       \and European Southern Observatory, Garching, Germany 
       \and California Institute of Technology, Pasadena, USA 
   \and Max-Planck-Institut f\"ur extraterrestrische Physik, Garching, Germany
   \and Osservatorio Astronomico, Bologna, Italy}
\maketitle

\begin{abstract}
The ROSAT Deep Survey in the {\it Lockman Hole} is the most sensitive X--ray
survey performed to date, encompassing an exposure time of 200 ksec with
the PSPC and 1.2 Msec with the HRI.
The source counts reach a density of $\sim1000~{\rm deg}^{-2}$
at a limiting flux of $\sim10^{-15}\erg$. At this level 70--80\%
of the 0.5--2 keV X--ray background is resolved into discrete sources.
Because of the excellent HRI positions, 83 X--ray sources with
fluxes (0.5--2 keV) above $ 1.2 \times 10^{-15}\erg$ could
be optically identified so far utilizing deep optical CCD images, NIR
photometry 
and Keck spectroscopy. Only 11 sources above this flux limit remain 
unidentified. The majority of objects turned out to be active
galactic nuclei (AGN) with minority contributions of clusters of galaxies,
stars and some individual galaxies.
These deep pencil beam data together with a number of shallower ROSAT
surveys define the source counts over six orders of magnitude in flux and
provide a unique tool of unprecedented quality to study the
cosmological evolution of AGN, which can easily provide the bulk of the
extragalactic X--ray background and could give an important contribution to
the total background light in the universe.
\end{abstract}

\section{Introduction}
Deep X--ray surveys using ROSAT, ASCA and BeppoSAX have resolved a
significant fraction of the cosmic X--ray background (XRB) into discrete
sources (Hasinger et al. 1998a, hereafter paper I) and optical 
identifications 
(Schmidt et al. 1998, hereafter paper II) are demonstrating that the XRB is 
largely due to accretion onto massive black holes, integrated over 
cosmic time. The deep soft X--ray surveys have detected a larger surface 
density of AGN than in any other waveband and find significant evolution 
in the space density of high--luminosity AGN contrary to the pure 
luminosity evolution which was the standard assumption so far
(Hasinger 1998). However, considerable
uncertainties still exist for the evolution of low-luminosity AGN
(Schmidt et al. 1999; Miyaji, Hasinger \& Schmidt, 1999) and at redshifts 
above 3. 

The characteristic hard spectrum of the XRB can only be explained
if most AGN spectra are heavily absorbed (see e.g. Comastri et al. 
1995). Thus as much as 80-90\% of the 
light produced by accretion may be absorbed by gas and dust clouds, which 
according to recent models could reside in nuclear starburst regions 
that feed the AGN (Fabian et al. 1998). This scenario would have important 
consequences
for the current attempts to understand black hole and galaxy formation
and evolution: The absorbed AGN will suffer severe extinction and 
therefore, unlike classical QSOs, would not be prominent at optical 
wavelengths. If most of the 
accretion power is being absorbed by gas and dust, it will have to be 
reradiated in the FIR range and be redshifted into the sub--mm 
band. AGN could therefore contribute a substantial fraction to the 
recently discovered cosmic FIR/sub--mm background which has already partly 
been resolved by deep SCUBA surveys (Almaini et al. 1999). The 
AGN light therefore needs to be taken into 
account when studying the star formation history in the early universe.

In this review we give a progress report of our studies of the 
XRB with emphasis on the ROSAT ultradeep HRI survey (section 2) and 
its optical/NIR identification (section 3). 
Surveys at harder X--rays are presented in context with the population 
synthesis models in section 4 and the X--ray luminosity
function of galaxies and AGN, its cosmological evolution, and the   
contribution of AGN to the history of star formation in the early 
universe are discussed in section 5. A Hubble constant of $H_0=50$ km/s/Mpc
and a deceleration parameter $q_0=0.5$ are assumed throughout the paper.

\begin{figure}[t]
\centerline{\psfig{figure=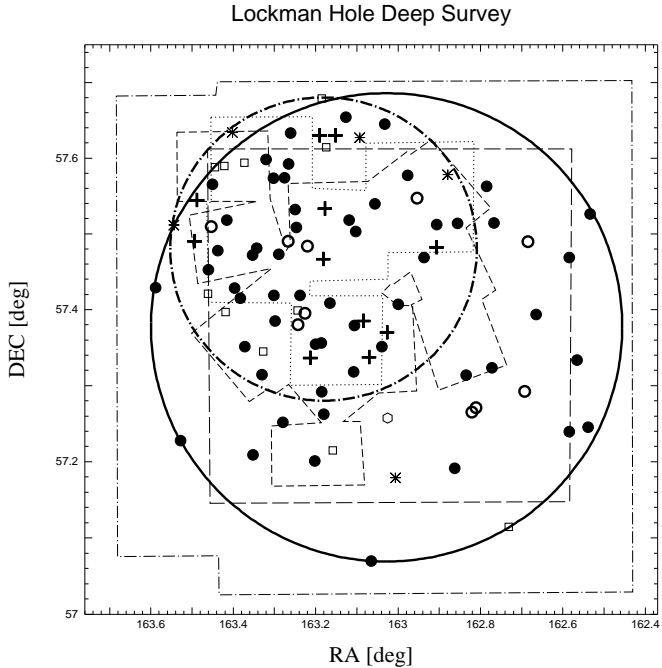,width=9.0cm}}
\caption{\small
Overview of X--ray surveys and optical/NIR coverage of the Lockman 
Hole. The thick solid circle gives the location of the PSPC deep survey 
with an 18.5 arcmin radius field. The thick dot--dashed 
circle shows the location of the HRI ultradeep survey 12 arcmin radius 
field. The thin dashed line outlines the Keck R--band images available
so far. Most of them have exposure times of 5 minutes, seeing 0.7" and 
limiting magnitudes about R=25.5 (see Lehmann et al. 1999). The thin dotted
line outlines the Calar Alto K--band images available so far (courtesy T. 
Stanke, M. McCaughrean). Exposure times are 
around 40 min each giving a limiting magnitude in K' of about 19.7. 
The long--dashed square indicates the area covered by  
UH 8K--images (courtesy G. Luppino) in V and I; 
seeing is about 0.7-1 arcsec, limiting magnitude around 24. 
The thin dash--dotted line shows the area covered by CCD
drift scans in V and I using the 4-shooter camera on the Palomar 200"
telescope (courtesy J. Gunn and D. Schneider). 
The field has also been mosaiqued with CCD 
exposures from the UH 88" in B and R (courtesy J. MacKenty and R. Burg). 
The symbols depict the location of the X--ray sources in our total sample.
Filled circles are spectroscopically identified broad-line AGN (ID 
classes a-c; see paper II). 
Open circles are narrow-line AGN (ID classes d and e; paper II). 
The open hexagon is one 
galaxy. Open squares are clusters of galaxies. Asterisks are stars.
Plus signs are the as yet unidentified sources.}
\label{SRV}
\end{figure}

\section{The ROSAT Ultradeep HRI Survey}

The ROSAT Deep Survey (RDS) project consists of pointed observations
with the ROSAT PSPC and HRI detectors in the direction of the 
Lockman Hole (paper I) and in the Marano field (Zamorani et al.
1999).
We have obtained a pencil beam survey in the Lockman Hole with the  
deepest X--ray observation ever performed. Images of 200 ksec observation 
time with the ROSAT PSPC (see Fig. \ref{SRV}) define the 
{\it Deep PSPC Survey} (Hasinger et al. 1993), and
exposures
totalling 1.2 Msec in a smaller area (Fig. \ref{SRV}) are the basis for the 
{\it Ultradeep HRI Survey} (paper I). The ultradeep ROSAT HRI 
survey now reaches
a surface density of $\sim 1000~{\rm sources~deg}^{-2}$ at a flux of
$10^{-15}\erg$ and 70--80\% 
of the soft X--ray background 
has been resolved into discrete sources. The fluctuation analysis
of the PSPC survey resolved about 85--100\% (Hasinger et al. 1993).

The Lockman Hole is also target for  
other deep multifrequency surveys.
In the optical band, a mosaique of UH 88" CCD images (B, R), Keck R--band 
CCD images, Palomar 5m drift scans (V, I) and UH 8K images (V, I) have been
obtained, which form the basis of our spectroscopic follow-up
identifications (paper II, see also Fig. \ref{SRV}). 
Deep VLA radio mosaic observations (deRuiter et al. 1997) and, recently, 
deep ASCA (Ogasaka et al. 1998) and BeppoSAX (Giommi 1999) hard X--ray 
images have been acquired in the $0.3~deg^2$ survey region. 
The field will soon be surveyed by the Chandra X--ray Observatory (AXAF)
and by XMM at hard X--rays in GT and PV-time, respectively.
The
Lockman Hole was also covered by a deep and medium-deep $7\mu$ and
$15\mu$ mid-IR survey with ISOCAM (Elbaz et al. 1998) and a $90\mu$ far--IR 
survey with ISOPHOT (Kawara et al. 1998) and is targeted in ongoing SCUBA 
observations.
Finally, the Lockman Hole is one of the CADIS fields (Thommes 
et al. 1998)
and a deep K--band survey with the Omega--Prime camera at Calar Alto
has been started (see Fig. \ref{SRV}). 

\section{Optical/NIR identifications}

Optical counterparts of the weakest X--ray sources are very faint 
($R>24$) and require good, unconfused X--ray positions and high--quality
optical spectra. For the Lockman Hole the optical
follow--up spectroscopy could largely be done with
long--slit and multi--slit spectroscopy at the Keck telescope. 
A catalogue of spectroscopic optical counterparts for a complete sample 
of 50 ROSAT PSPC sources in the Lockman Hole with 0.5--2 keV fluxes brighter
than $1.1 \times 10^{-14}\erg$ in a solid angle of 
$0.299~deg^{-2}$ and fluxes brighter than
$5.5 \times 10^{-15}\erg$ in a solid angle of 
$0.136~deg^{-2}$ has been published in paper II.
The large majority ($>80\%$) of 
the X--ray sources in the deep PSPC survey turned out to be AGN.
Most of those are QSOs and Sy1 galaxies with at least one broad emission
line in their optical spectra. A non--negligible fraction ($\sim 16\%$)
shows only narrow emission lines in their spectra. We interpret
these as type 2 AGN because of the presence of high excitation
[NeV] emission lines and/or high X--ray luminosity ($L_X>10^{43}\ergs$). 
In the published catalogue of the 50 X--ray sources in the deep PSPC survey,
there are four unidentified objects. In the meantime one of
these four sources (\#36) has been identified as an AGN at z=1.52
(Lehmann et al. 1999).
The status of other unidentified sources are discussed in the following
together with the sample of X--ray sources from the ultradeep HRI survey
with fluxes fainter than $5.5 \times 10^{-15}\erg$. 
Among the latter objects is the highest redshift X--ray selected QSO at 
z=4.45 (Schneider et al. 1998).

\begin{figure*}
\begin{center}

\begin{minipage}{160mm}
\begin{minipage}{80mm}
\psfig{figure=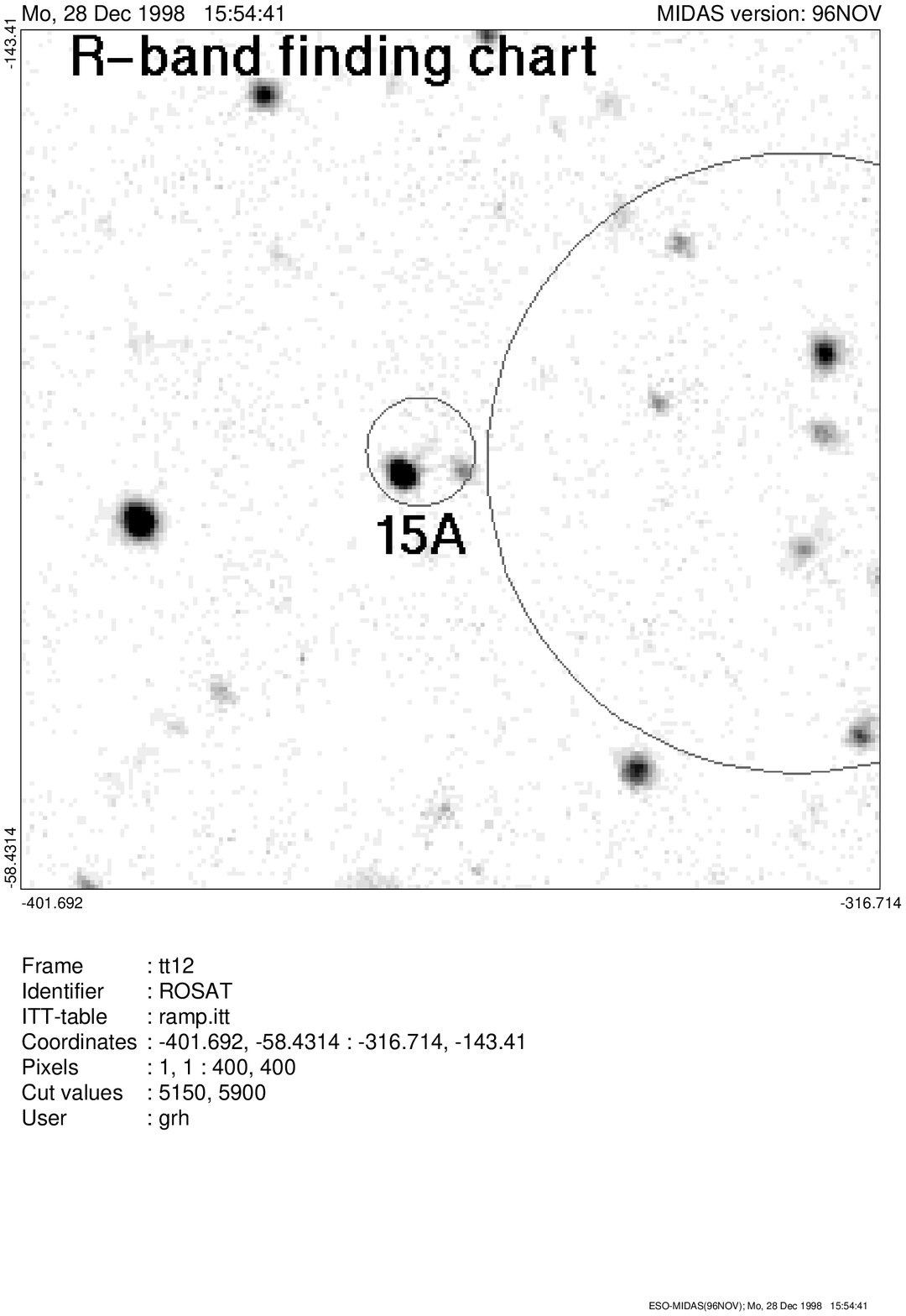,bbllx=86pt,bblly=300pt,bburx=530pt,bbury=745pt,height=80mm,width=80mm,clip=}
\end{minipage}
\begin{minipage}{80mm}
\psfig{figure=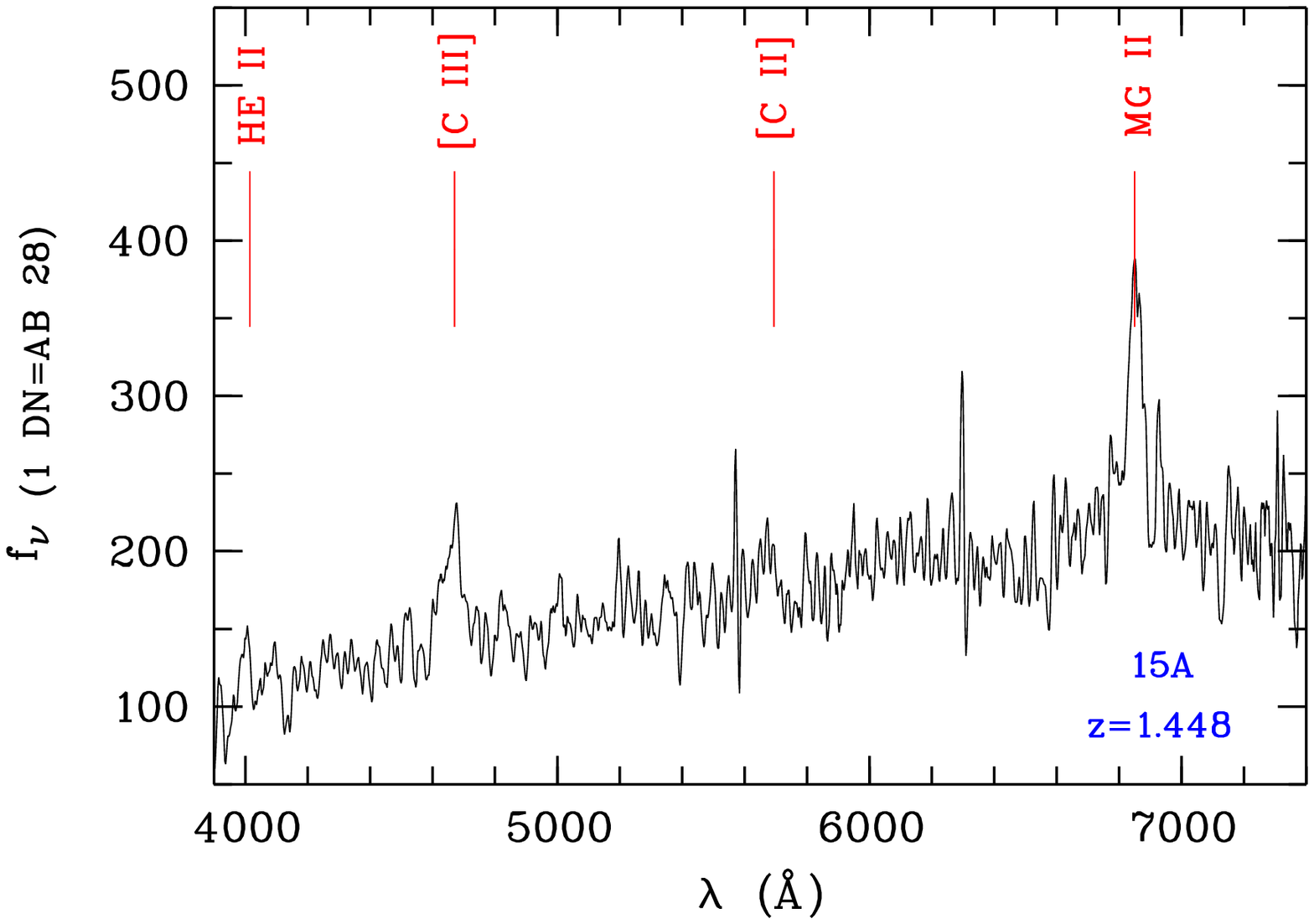,width=80mm,height=80mm,angle=-90,clip=}
\end{minipage}

\end{minipage}
\begin{minipage}{160mm}
\begin{minipage}{80mm}
\psfig{figure=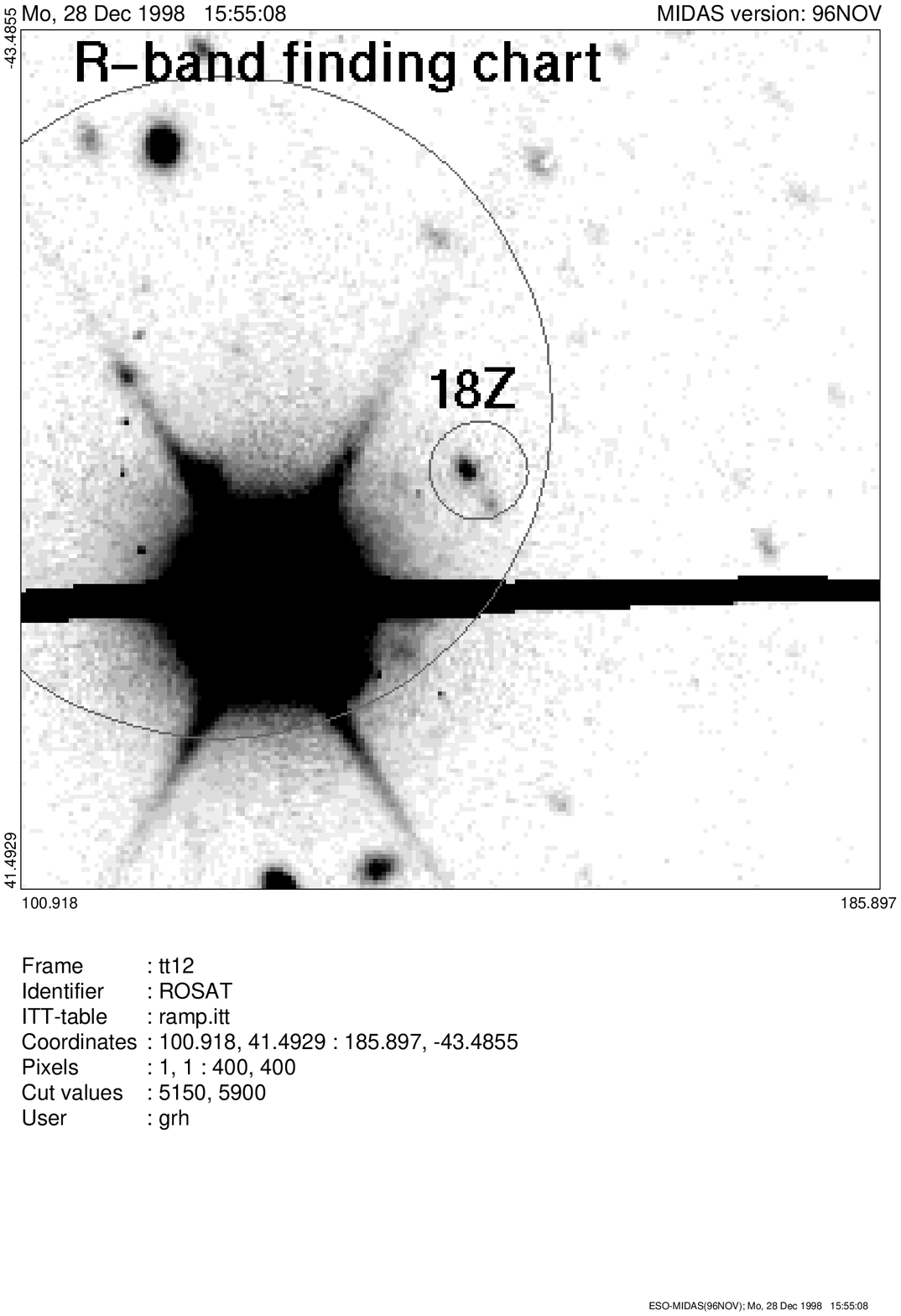,bbllx=86pt,bblly=300pt,bburx=530pt,bbury=745pt,height=80mm,width=80mm,clip=}
\end{minipage}
\begin{minipage}{80mm}
\psfig{figure=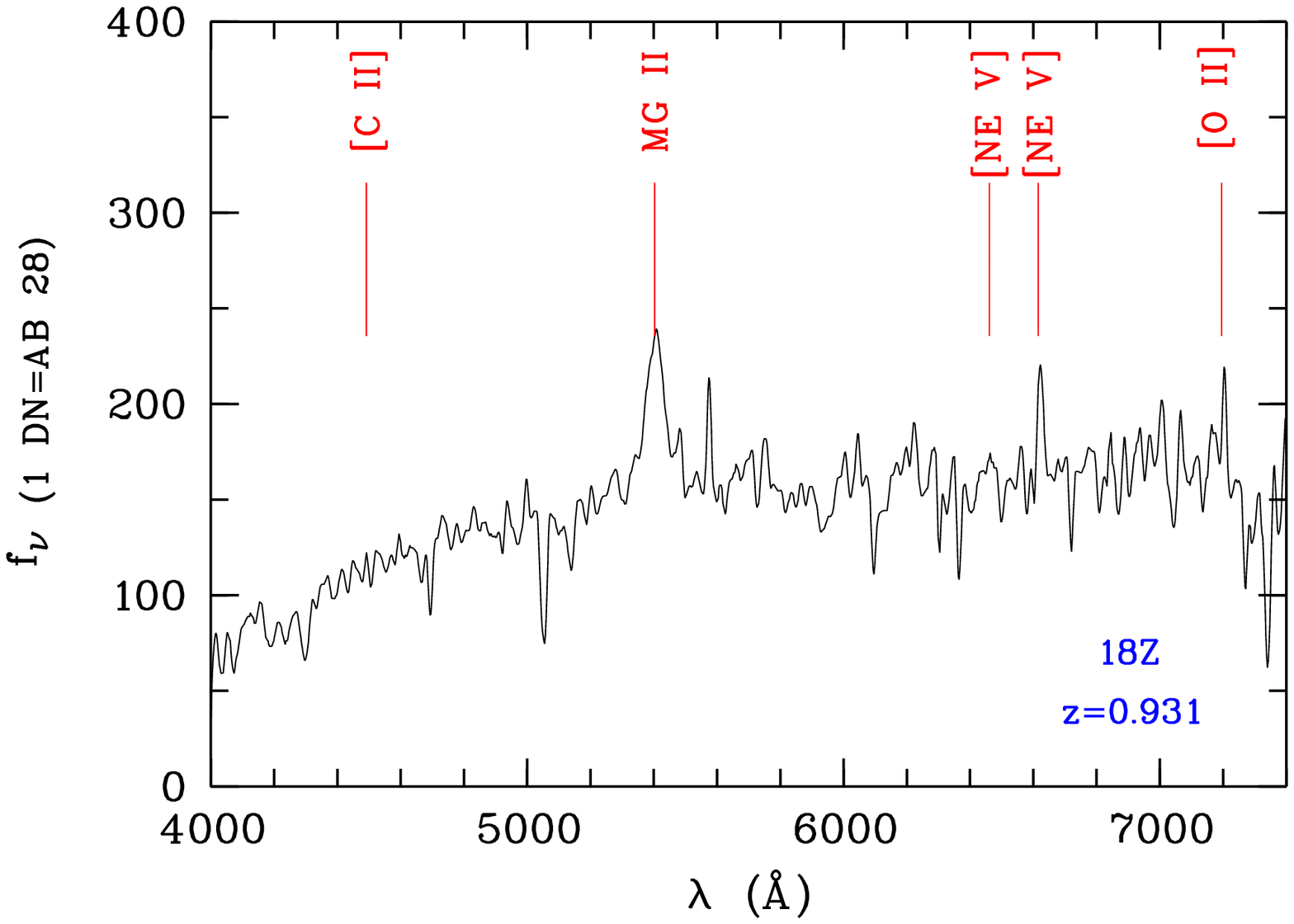,width=80mm,height=80mm,angle=-90,clip=}
\end{minipage}
\end{minipage}

\begin{minipage}{160mm}
\begin{minipage}{80mm}
\psfig{figure=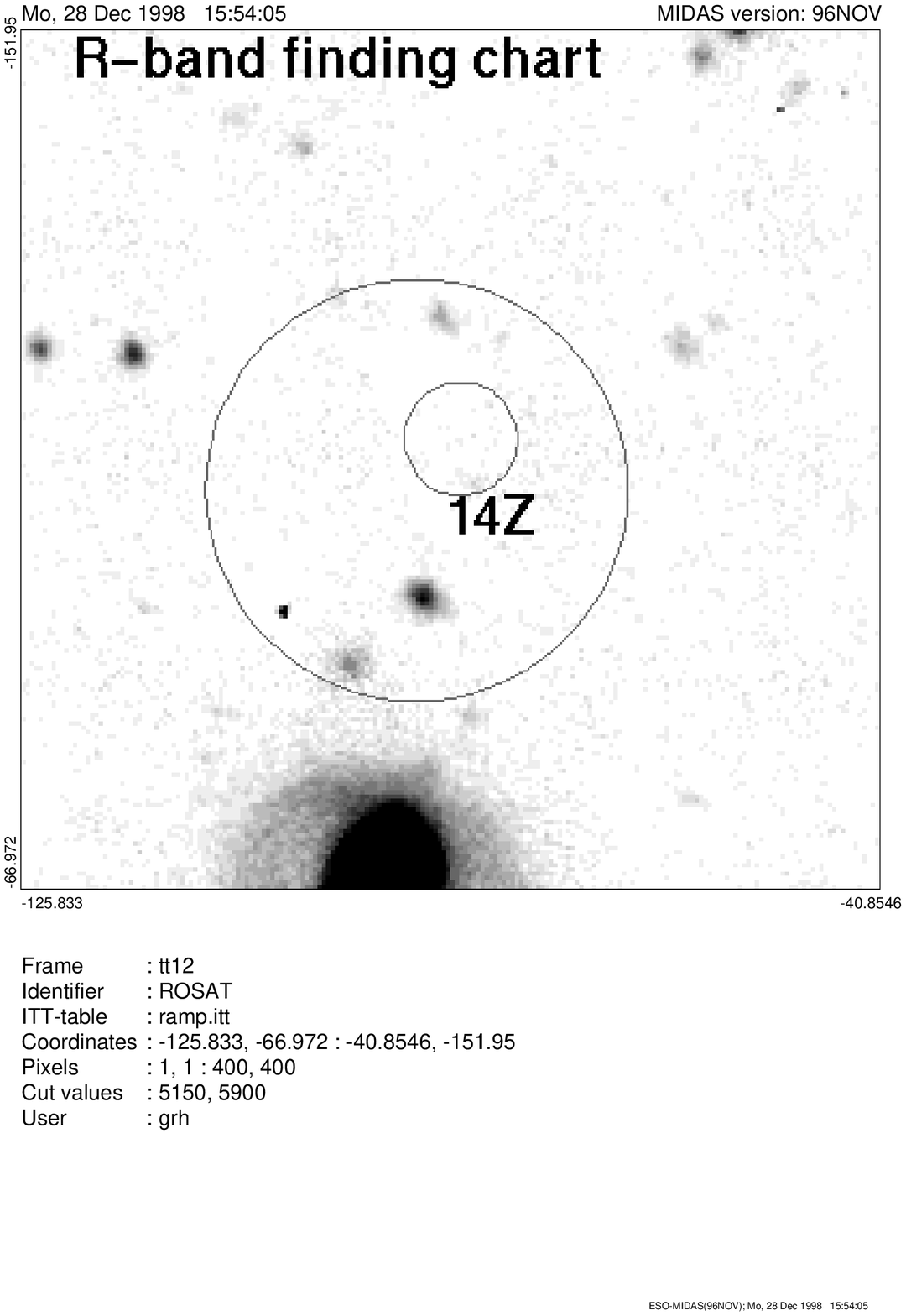,bbllx=86pt,bblly=300pt,bburx=530pt,bbury=745pt,height=80mm,width=80mm,clip=}
\end{minipage}
\begin{minipage}{80mm}
\psfig{figure=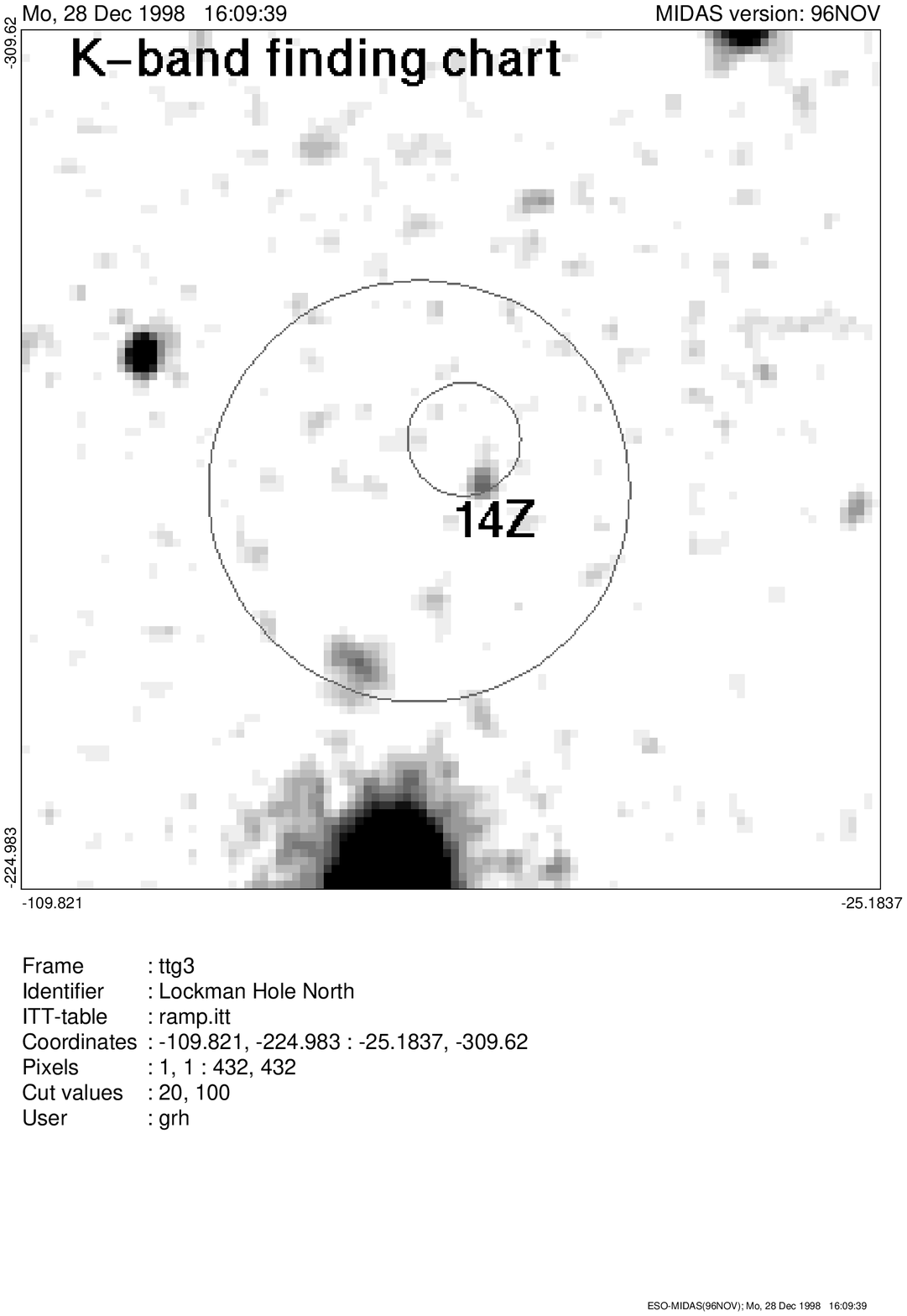,bbllx=86pt,bblly=300pt,bburx=530pt,bbury=745pt,height=80mm,width=80mm,clip=}
\end{minipage}
\end{minipage}

\caption{\small
Finding charts ($85" \times 85"$) and optical spectra of selected 
X--ray sources in the ROSAT ultradeep HRI survey. 
The small circles and large error circles are from HRI and PSPC, respectively.}
\label{OPT}
\end{center}
\end{figure*}

\subsection{Sample selection}

There are systematic differences between the
deep PSPC and the ultradeep HRI pointing in the Lockman Hole. The HRI 
pointing center is shifted with respect to the PSPC center by 
about 9 arcmin in North-East direction (paper I, see Fig. \ref{SRV}). 
The conversion
between count rates observed in the PSPC hard band (PI channels 51-201)
to 0.5--2 keV fluxes is straightforward because of the similar bandpasses,
while the HRI detector is sensitive in the 0.1--2 keV range leading
to a substantial model--dependence in the count rate to flux conversion.  
Time variability 
of X--ray sources can also lead to different fluxes for the same source 
in different exposures (see e.g. paper II). 
In order to nevertheless construct a well defined, reliable set 
of faint X--ray sources combining both the PSPC and the HRI data we
have used a sample definition different than that in paper I.  

\begin{figure}[t]
\centerline{\psfig{figure=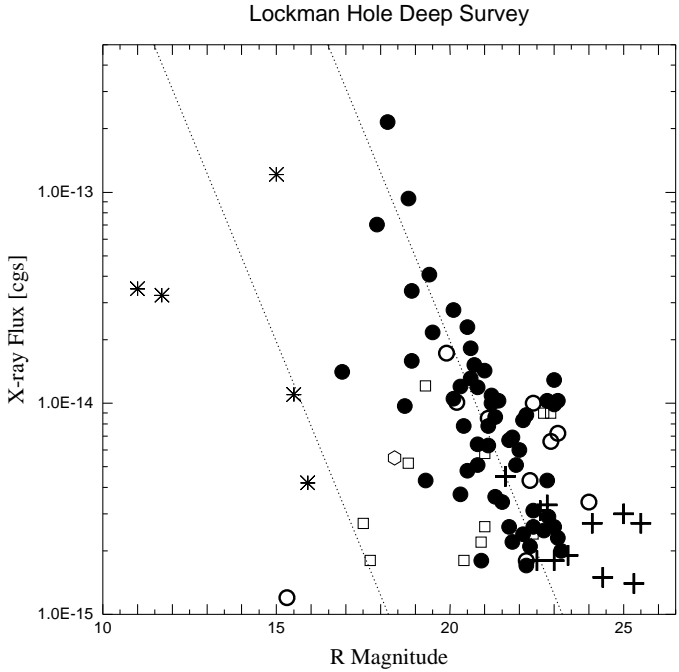,width=9.0cm}}
\caption{\small
Correlation between X--ray flux and optical R magnitude
for all objects in the Lockman Hole survey. Symbols are the same as 
in figure \ref{SRV}. 
For the unidentified sources we show
the brightest optical counterpart in the 90\% error circle.}
\label{FXO}
\end{figure}

First we have selected all HRI sources with off--axis angle less than 
12 arcmin from the HRI pointing center. Their 0.5--2 keV X--ray fluxes were
determined from the HRI count rates and a renormalized count--to--flux 
conversion factor using all sources detected jointly in the HRI and
the PSPC. Thus the epoch and fluxes of this part of the survey are defined 
by the HRI pointing. The approximate flux limit of the HRI survey is
$1.2 \times 10^{-15}\erg$ (see paper I) and the solid angle 
$0.126~{\rm deg}^{-2}$.
In addition, we use the deep PSPC sources which do not fall into
the HRI sample and have PSPC off--axis angles between 12.5 and 18.5 arcmin
and PSPC fluxes larger than $9.6 \times 10^{-15}\erg$ 
(solid angle $0.071~deg^{-2}$) or PSPC off--axis angles smaller than 
12.5 arcmin and PSPC fluxes larger than 
$5.5 \times 10^{-15}\erg$ (solid angle $0.119~deg^{-2}$).
This latter part of the sample is statistically independent from the
HRI sample and defined by the epoch of the PSPC observations. 
It is very similar to that part of the sample in papers I and II, which 
is not covered by the HRI pointing. 
The total number of sources in the sample is 94 (68 HRI and 26 PSPC).
The new sample definition results in some changes compared to the catalogue 
of the 50 brightest sources presented in paper I and II. A few sources are 
lost completely (e.g. the previously unidentified source \#116), others 
have a higher flux because of time variability.

\begin{figure}[t]
\centerline{\psfig{figure=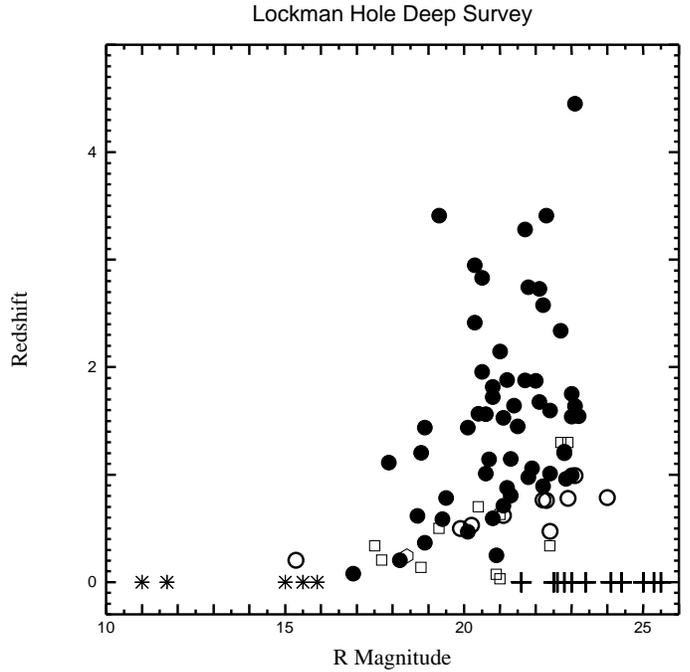,width=9.0cm}}
\caption{\small
Correlation between optical R magnitude and redshift
for all objects in the Lockman Hole survey. 
Symbols as in Fig. \ref{SRV}. Unidentified sources (crosses)
are shown at redshift zero.}
\label{RXZ}
\end{figure}

The importance of an excellent X--ray position accuracy in order to
avoid confusion and to obtain unique identifications of extremely
faint sources has been emphasized in paper I and II. 
Figure \ref{OPT} shows three examples of optical finding charts, 
based on deep Keck R--band exposures, where the very good HRI positions 
(90\% error radius 2--5 arcsec) pointed out unique optical counterparts
which would have been very hard to identify from the PSPC positions
alone. The first example (\#15) is a case where a z=1.45 QSO, pinpointed 
uniquely by the HRI, lies outside the relatively large PSPC 
90\%--error circle. The second source (\#18) was originally  
misidentified with the bright star in the PSPC error circle, until the
HRI pointed out a z=0.93 QSO only a few arcsec from the star.
The third case is ROSAT \#14 where no optical counterpart is found in
the HRI error circle to $R<24.5$. A 45min K'--band exposure with the 
OMEGA-Prime camera on the Calar Alto 3.5m telescope was obtained 
in December 1997. Interestingly, the X--ray error circle contains
a K'=19.5 infrared counterpart associated with an R=25 optical object
and a very red colour of $R-K'=5.5$. We do not yet have a
spectroscopic identification of this object, but assume that it is 
a high--redshift and/or obscured AGN. 

In the Lockman Hole Deep survey most
optical counterparts have magnitudes in the range R=18--25.5.
With the excellent HRI positions typically
only one or two counterparts are within the X--ray error circle. In Fig.
\ref{FXO} the X--ray fluxes of our sources are plotted against the magnitude
of their optical counterparts or, in case of no identification, of the 
brightest optical candidate in the error box. The objects with 
spectroscopic identifications are marked with different symbols,
while the unidentified sources are shown with crosses. As shown
in this figure, we have a 100\% complete identification for 53 sources
with fluxes above $5 \times 10^{-15}\erg$ in this sample,
while at fainter fluxes 11 out of 41 sources remain unidentified.
The large majority of optical identifications even at these faint fluxes 
are broad-line AGN (filled circles)  
with X--ray/optical flux ratios typically scattering
around unity (right dotted line). This is in marked contrast to the 
findings of McHardy et al. (1998), who in their deep PSPC survey claim to find 
a majority of galaxies
(or narrow-line AGN) which have similar X--ray fluxes but on average lower
$f_X/f_{opt}$ ratios. This indicates a substantial fraction of 
misidentifications in the latter study. 

Figure \ref{RXZ} shows a correlation between optical magnitude and
redshift for all X--ray sources. For the purpose of deriving luminosity
functions, it is important to note, 
that about 20\% of our AGN have $z>2$ and about 40\% $z>1.5$. Clusters 
of galaxies and narrow-line AGN are typically found at redshifts below 1. 

\begin{figure}[t]
\centerline{\psfig{figure=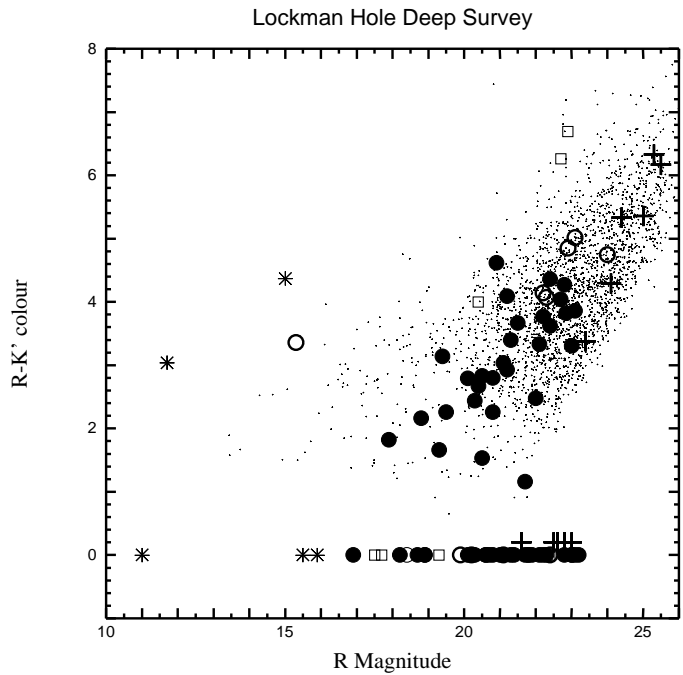,width=9.0cm}}
\caption{\small
Colour-magnitude diagram between R magnitude and $R-K'$
colour for objects in the Lockman Hole survey. 
Small dots are all sources detected in the 
Lockman Hole K--band survey. 
Symbols of X--ray sources as in Fig. \ref{SRV}. Those X-ray sources
not covered by the K survey are plotted at $R-K'=0$ or 0.2.
The optically faintest unidentified sources have and the galaxies associated
with a high--redshift cluster (see below) have red $R-K'$ colours.}
\label{RMK}
\end{figure}

\subsection{NIR photometry}

Optical spectroscopy from Keck becomes exceedingly time consuming
for objects with $R>23.5$, which is roughly our current spectroscopic
limit. A few of the unidentified sources in Fig. \ref{FXO} are brighter
than this limit. For some of those we just were not able to take 
spectra yet and some show noisy spectra without obvious features. 
Five unidentified sources have very  
faint optical counterparts ($R>23.5$). For these objects we were able
to obtain NIR images down to $K'\sim$19-20. Because of the large
field of view of Omega-Prime ($42~arcmin^2$), these images cover almost 
half of the ultradeep HRI survey field. We are therefore able to
derive R--K' colours for almost half of our sample. Figure \ref{RMK}
shows a colour-magnitude diagram for all objects detected in the 
K'--images, including the X--ray sources in the K--band survey.    
This figure clearly shows, that all X--ray counterparts with optical
magnitudes fainter than R=23.5 have red colours ($R-K'>4$).
The identified X--ray sources with such red colours are either
clusters of galaxies at redshifts above 1 (see below) or low-luminosity,
absorbed AGN at various redshifts. We can therefore safely
assume that the faint unidentified sources are part of the same 
population as the identified sources. A similar group of red 
X-ray sources has been found by Newsam et al. (1997).

\begin{figure}[htp]
\centerline{\psfig{figure=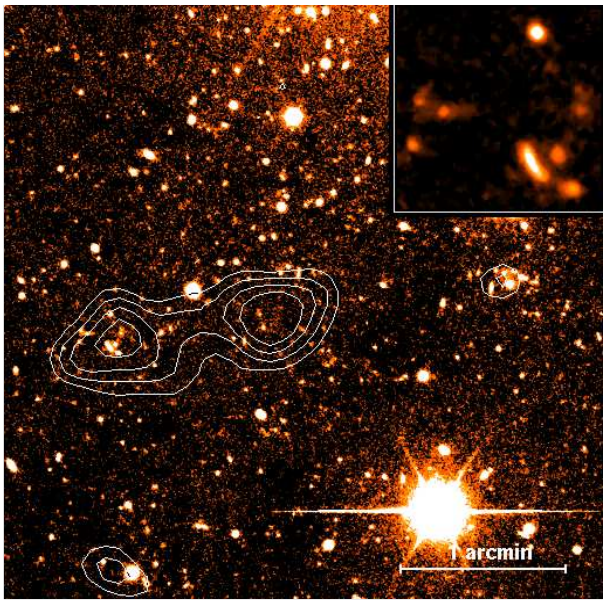,width=8.0cm,clip=}}
\caption[ ]{\small ROSAT ultradeep HRI pointing X--ray contours of a region
about $3\times3 {\rm arcmin}^2$ in the Lockman Hole, superposed on a 5 min
Keck R band exposure. North is up and East is to the left. Three distinct 
X--ray sources are detected in this image: the extended source 
RX J105343+5735 with its eastern and western lobe, a cluster of galaxies
at z=0.7 to the West and a QSO at z=2.572 to the South. The insert in the 
upper right is a $20 \times 20~arcsec^2$ zoomed cutout of the image
close to the center of the eastern lobe. A gravitationally lensed arc at
a redshift of 2.570 is detected there.}
\label{CLU}
\end{figure}

\subsection{Clusters of galaxies}

The second most abundant class of faint X--ray sources 
in our sample are clusters 
and groups of galaxies (open squares in Fig. \ref{FXO}).
Their surface density in our survey is $63\pm22~{\rm deg}^-2$, 
consistent with an extrapolation of the no-evolution 
log(N)--log(S) found in the ROSAT Deep Cluster 
Survey by Rosati (1998) and their redshifts are typically below 0.7. 
A substantial
fraction of the clusters and groups appear extended in the ultradeep
HRI image (see paper II). 

\begin{figure}[htp]
\centerline{\psfig{figure=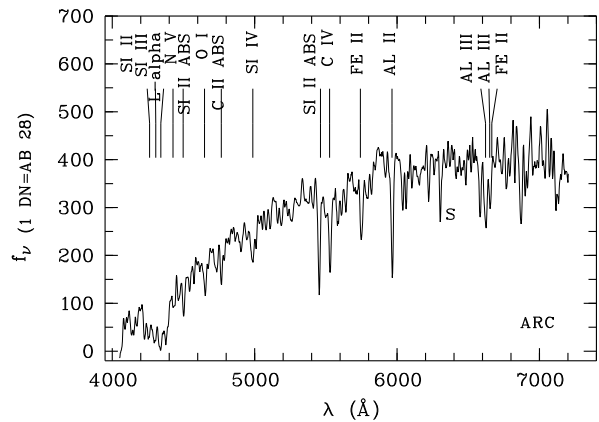,width=8.0cm,angle=-90}}
\caption[ ]{\small Keck LRIS spectrum of the arc in RX J105343+5735.
The wavelengths of prominent UV metal absorption
lines, redshifted to z=2.570 are indicated. ``S'' denotes the position of a
strong sky line. The spectrum is typical for a high--redshift starforming
galaxy.}
\label{ARC}
\end{figure}

The brightest extended X--ray source in the Lockman
Hole (RX J105343+5735) shows a peculiar elongated double-lobed structure
(see Fig. \ref{CLU}).
The angular size of the X--ray source is $1.7 \times 0.7~{\rm arcmin}^2$,
its X--ray flux $2 \times 10^{-14}\erg$.
R--band optical imaging from the Keck telescope revealed only a marginal 
excess of galaxies brighter than R=24.5. The brightest
galaxy close to the center of the eastern emission peak turned out
to be a gravitationally lensed arc at z=2.570 (see Fig. \ref{ARC}), 
suggesting that the X--ray emitting object is most likely a cluster of 
galaxies. This is the first
detection of a gravitational arc which is optically brighter than
any of the components of the lens.  X--ray and 
optical data on this cluster have been published in a recent 
paper (Hasinger et al. 1998b), where it was argued that
RX J105343+5735 is a moderately luminous cluster at a redshift
around unity,
based on a comparison of lensing surface mass density, X--ray
luminosity, morphology and galaxy magnitudes with
a large sample of clusters of known distance.
In the meantime, the cluster was covered by one of the K--band images
(see above), which clearly shows relatively bright, very red
($R-K'>6$)
cluster galaxies at the X--ray centroids of the two lobes (see also fig
\ref{RMK}). A very high redshift ($\sim 1.3$) is thus expected for this 
object.

\section{Hard X--ray surveys}

The X--ray background has a significantly harder spectrum than that of
the sources resolved in the soft band. This led to the assumption that
a large fraction of the background flux is due to obscured AGN, as
originally proposed by Setti and Woltjer (1989). Models following the
unified AGN schemes, assuming an appropriate mixture of absorbed and 
unabsorbed
AGN spectra folded with cosmological AGN evolution models, can quite 
successfully
explain the shape of the background spectrum over the whole X--ray band as
well as a number of other observational constraints (e.g. Comastri 
et al. 1995). The distribution of absorption column densities among
different types of AGN is one of the major uncertainties in this kind 
of modelling (see below). Observationally, it has so far only been possible 
to derive this for local Seyfert galaxies selected at hard X--rays (see e.g.
Schartel et al. 1997) or from optical emission lines (Maiolino et al. 1999).
The standard X--ray background population synthesis models (Matt \& Fabian 
1994; Madau et al. 1994; Comastri et al. 1995) assume that the absorption 
distribution is independent of X--ray luminosity and redshift. 

An immediate consequence of these types of models is that the radiation
produced by accretion processes in AGN emerges completely 
unabsorbed only at energies well above 10 keV, thus producing the observed 
maximum of the XRB energy density at $\sim 30$ keV . Comparison between
the background energy density at 30 keV and at 1 keV leads to the 
suggestion that most (80--90\%) of the accretion power in the universe
might be absorbed, implying a very large solid angle of the obscuring 
material as seen from the central source. Fabian et al. (1998) suggest that
circumnuclear starburst regions are responsible for the large covering
factor. They may be both triggering and obscuring most of the nuclear
activity. Recently Fabian and Iwasawa (1999) have shown that these  
AGN background synthesis models can 
explain the mass distribution of dark remnant black holes in the 
centers of nearby galaxies by conventional accretion which is 
largely hidden by obscuration.  

The existing background synthesis models predict that a large part of the hard 
X--ray background comes from significantly
absorbed objects, which are almost absent in the soft band, even
at the faintest ROSAT limit. As a consequence, a significant test for these
models would be the comparison of their predictions with the results
of optical identifications of a complete sample of sources
selected at faint fluxes in the hard X--ray band.

Because of the technological challenge for X--ray imaging above
2 keV, hard X--ray surveys are just becoming available now.
A large sky survey (LSS) with a limiting flux of 
$ 10^{-13}\erg$ (2--10 keV) near the North Galactic Pole
was performed with the ASCA satellite (Ueda et al. 1998).
Recently, ASCA data have been used by Cagnoni et al. (1998) 
to derive the 2--10 keV log(N)--log(S) down to fluxes slightly below 
$10^{-13}\erg$. 
The deepest ASCA surveys (Georgantopoulos et al. 1997; Ogasaka et al. 1998)
resolve source counts down to
2--10 keV fluxes of $5\times10^{-14}\erg$. At 
surface densities of typically $100~{\rm sources~deg}^{-2}$ these surveys 
are heavily confused due to ASCA's limited angular resolution. 
An analysis of the spatial fluctuations in deep ASCA images (Gendreau et al.
1997) probes the 2--10 keV X--ray source counts down to a flux limit of
$2\times 10^{-14}\erg$, resolving about 35\% of the 
extragalactic 2--10 keV X--ray background.

A new High Energy Large Area Survey (HELLAS) has been started with 
BeppoSAX in the 5--10 keV band, which is particularly well suited
for this instrument because of the relatively large throughput at
high energies and a significantly sharper point spread function 
compared with ASCA. A surface density of $\sim20~{\rm sources~deg}^{-2}$
is reported (Fiore et al. 1998) at a 5--10 keV flux limit of 
$5\times 10^{-14}\erg$, indicating that 30--40\% of 
the background in this energy band has already been resolved.   

The hard X--ray log(N)--log(S) data so far are in good agreement with the
predictions of the population synthesis models (Comastri et al. 1999;
Miyaji, Hasinger \& Schmidt 1999).
Following these models, a large fraction of X--ray sources at faint
fluxes should be substantially
absorbed and therefore their counterparts are expected to have
optical spectra typical of Seyfert 2 galaxies. Programs to optically 
identify the sources from these hard surveys have already started and
indeed find typically AGN counterparts (Boyle et al. 1998; Akiyama et 
al. 1998; Fiore et al. 1999), but the 
large positional uncertainty together with the relatively faint optical 
counterparts slows down progress so that sample sizes are still small. 

Deep hard surveys with ASCA and BeppoSAX have also been taken 
in the Lockman Hole, where due to the existence of the ROSAT 
HRI data a cross--identification between the soft and hard X--ray data
is readily available. Details of the ASCA and BeppoSAX surveys will be 
presented elsewhere (Ishisaki et al. 1998; Giommi 1999). A 
somewhat surprising result is that almost all hard X--ray sources in the 
ASCA and BeppoSAX images of the Lockman Hole have relatively bright soft X--ray
counterparts. This appears to be inconsistent with the simple XRB population 
synthesis models, which would predict a substantial fraction of hard sources 
not detectable in the soft band. We may see here effects that have 
been neglected in the population synthesis models, like e.g. partial 
obscuration or unabsorbed soft spectral components in heavily obscured AGN. 
Another possible effect 
would be a dependence of the obscuration distribution on luminosity and/or 
redshift. Despite occasional discoveries of absorbed (type 1.5-2) QSOs 
(Almaini et al. 1995; Ohta et al. 1997; Zamorani et al. 1999), 
there is observational evidence that high luminosity and/or high 
redshift AGN are on the average much less absorbed than local Seyfert 
galaxies (see e.g. Fig. \ref{RXZ}; Halpern et al. 1998; Miyaji, Hasinger 
\& Schmidt, 1999). 

Detailed information about the absorption distribution of the AGN population
as a function of luminosity and redshift is a necessary ingredient
to derive the AGN X--ray luminosity function (XLF) and its cosmological 
evolution, which in turn is input into the population synthesis models for 
the X--ray background. So far, the derivations of the AGN XLF have largely 
ignored the effects of X--ray absorption (see below). 
First global, simultaneous fits of the XLF, X--ray background spectrum and 
absorption distribution have just been performed (Schmidt
et al. 1999; Miyaji, Hasinger \& Schmidt, 1999a), but are still quite
uncertain because of the large number of parameters and the possible 
hidden correlations involved. 
Upcoming deep surveys with the Chandra observatory (AXAF) and XMM 
with very high sensitivity and good positional accuracy in the hard band 
together with optical identifications from the VLT and the Keck telescopes
are expected to yield a solid statistical basis to disentangle these 
various effects and lead to a new, unambiguous population synthesis for the 
X--ray background.

\section{AGN cosmological evolution}

As discussed above, information about the cosmological evolution of the AGN 
population is a crucial input into the background synthesis models, but it can 
not be obtained without taking into account the AGN absorption 
distribution. However, the AGN X--ray luminosity function in the 0.5--2 keV
band has so far mainly been derived ignoring the effects of absorption. 
First attempts to study AGN cosmological evolution
from the Einstein Medium Sensitivity Survey (EMSS; Della Ceca et al. 1992)
or from a combination of medium deep ROSAT fields with the EMSS (Boyle et 
al. 1994) determined the local AGN XLF which has the shape of a broken
power law. Their data
are consistent with pure luminosity evolution proportional to $(1+z)^{2.7}$
up to a redshift $z_{max} \approx 1.5$, similar to what was found previously
in the optical range. This result has been confirmed and improved  
by more extensive or deeper studies of the AGN XLF, e.g. the RIXOS
project (Page et al. 1996) or the UK deep survey project (Jones et al. 1997).
However, these studies were hampered by the uncertain crosscalibration 
between Einstein and ROSAT.

In the meantime optical identifications of a large number of ROSAT X--ray 
surveys 
at various flux limits and solid angle coverage have been completed, so that
a new AGN soft X--ray luminosity function could be determined, based on
ROSAT surveys alone (Hasinger 1998; Miyaji, Hasinger \& Schmidt, 1998+1999b; 
Schmidt et al. 1999). The log(N)--log(S) function
of the overall sample covers six orders of magnitude in flux and agrees
within $\sim 10\%$ in the overlapping flux regions between different surveys
(Hasinger 1998). 

\begin{figure}[t]
\centerline{\psfig{figure=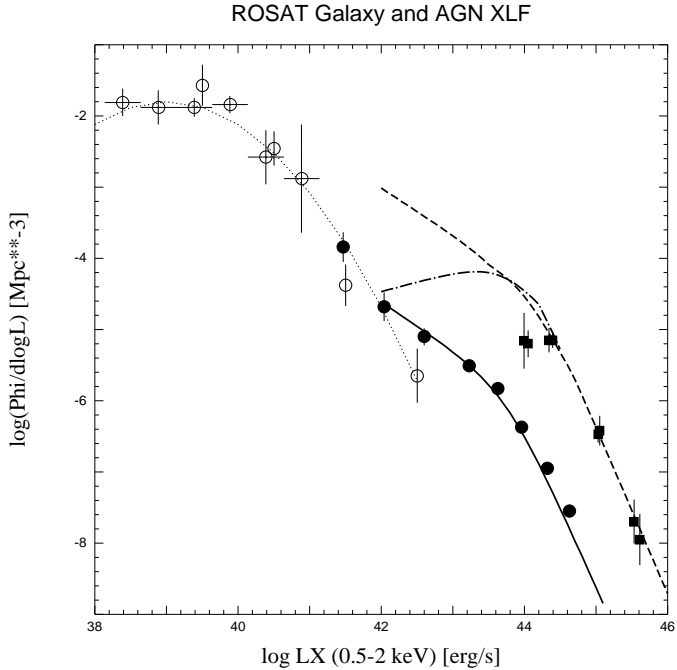,width=9.0cm}}
\caption{\small
X--ray (0.5--2 keV) Luminosity functions for AGN and galaxies from
ROSAT surveys. The local galaxy luminosity function (open circles
and dotted line) has been
derived by Hasinger (1998) from the ROSAT Bright Survey 
(Fischer et al. 1998; Schwope et al. 1999) and from a volume-limited sample 
of local galaxies (Schmidt et al. 1996).  The
AGN luminosity function (filled circles, from Miyaji et al. 1999b) is shown 
only for nearby objects ($z<0.2$) and very distant objects ($z>1.6$).
Two different luminosity--dependent density evolution models have been
fit to the data, one which is close to a pure density 
evolution model (LDDE2, Miyaji et al. 1999b; dashed line) and one 
where evolution slows down substantially for low luminosities (LDDE1, 
Miyaji, Hasinger \& Schmidt, 1998; dash--dotted line). Both models
are consistent with all available constraints; their predictions 
for the density of low luminositiy AGN, however, diverge by 
more than one order of magnitude.}
\label{XLA}
\end{figure}

Contrary to the previous findings, the new XLF is not
consistent with pure luminosity evolution. For the
first time we see evidence for strong cosmological evolution of the
space density of low-luminosity AGN (e.g. Seyfert galaxy) XLF out to a
redshift 1--2, incompatible with pure luminosity evolution.
Pure density evolution proportional to $\sim (1+z)^{5}$ provided a 
reasonable fit
to the ROSAT data (Hasinger 1998), but overpredicts the total X--ray 
background, when extrapolated to lower luminosities.  
Therefore more complicated evolution models have to be taken into account.
The latest treatments (Schmidt et al. 1999; Miyaji, Hasinger \& Schmidt, 
1998+1999b)
agree that luminosity--dependent density evolution (LDDE) models, where the 
rate of density evolution is a function of luminosity, can match 
all constraints. This evolutionary behaviour is similar to the most recently 
determined optical QSO evolution (Wisotzki 1998). 

Figure \ref{XLA} compares the luminosity function for local and 
high--redshift AGN to the luminosity function of local normal and 
star bursting
galaxies. The low-redshift AGN XLF connects smoothly to the galaxy XLF at 
X--ray
luminosities of $L_X\approx10^{42}\ergs$. Around this luminosity there is 
some ambiguity about the relative contribution between the nuclear AGN light 
and diffuse galactic X--ray emission processes (see Lehmann et al. 1998). 
For clarity, 
measurements of the high--redshift AGN XLF are only shown for the two highest 
redshift shells ($1.6<z<2.3$ and $2.3<z<4.5$) from the data of Miyaji, 
Hasinger \& Schmidt (1998). 
The apparent deficiency in the lowest luminosity bins is most likely due to 
incompleteness. 
Two luminosity--dependent density evolution models 
are shown, which fit all observational constrains well: the LDDE1 model 
from Miyaji, Hasinger \& Schmidt (1999) (dash--dotted line), which is
similar to the LDDE model of Schmidt et al. (1999) has a rapid slow--down
of the density evolution below X--ray luminosities of $10^{44}\ergs$ and
produces $\sim60\%$ of the extragalactic 0.5--2 keV background.
The LDDE2 model (Miyaji et al. 1999) is not very 
much different from a pure density evolution model and produces 
$\sim90\%$ of the soft background. The constraints for the XLF of 
faint, high--redshift Seyfert galaxies, which can produce a significant 
fraction of the soft X--ray background and, depending on absorption 
properties, an even larger fraction of the hard X--ray background, are 
therefore still quite uncertain (the range is a factor of $\sim 25$ at 
$log L_X=42$). The LDDE2 model predicts a high--redshift AGN space density
which is close to that of local normal galaxies just above the break
of the luminosity function and to that of high--redshift galaxies
selected as U--dropouts (Pozzetti et al., 1998).
On the contrary, the LDDE1 model predicts a dearth of 
high--redshift Seyfert galaxies. A choice between these two possibilities 
will soon be possible with the even deeper X--ray surveys to be performed 
with the Chandra and XMM observatories.

\begin{figure}[t]
\centerline{\psfig{figure=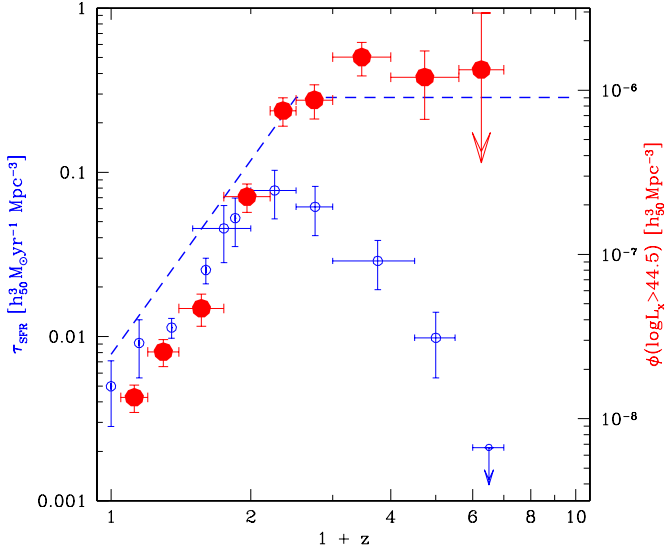,width=9.0cm}}
\caption{\small Cosmic star formation history $\tau_{SFR}$
(left Y-axis) compared to the space density $\phi$ of luminous X--ray 
selected QSOs (right Y-axis). Filled circles give the comoving number
density of ROSAT QSOs with $log~L_X>44.5\ergs$ (from Miyaji et al.
1998). Open circles give the optical/UV measurements of the star formation 
rate compiled by Blain et al. 1998. The dashed line indicates the simplest
star formation history model by Blain et al., which explains the whole
FIR/sub--mm background light by dusty star formation. Note the similarity 
between this model and the QSO space density.} 
\label{ZLF}
\end{figure}

\subsection{The space density of high--redshift QSO}

The X--ray data can also give important new information on the AGN 
evolution at very high redshifts and therefore on the epoch of black hole 
formation and the accretion history in the early universe. It is well
known from optical samples that the strong evolution of the space 
density of high luminosity QSOs slows down beyond a redshift of 
$\sim 1.5$ and that the space density decreases significantly beyond 
$z\approx2.7$ 
(see e.g. Schmidt, Schneider \& Gunn 1995). Different selection techniques 
have to be used below and
above a redshift of $\sim 2.2$ leading to possible systematic uncertainties
in the optical data. Radio--selected QSOs, however, confirm the decline
at high redshift and indicate that it is apparently not due to an increase
in dust obscuration (Shaver et al. 1998). The ROSAT sample of QSOs 
allows now for the first time to determine the space density of X--ray selected
AGN in the whole range  of $0<z<5$ with one technique. Figure \ref{ZLF}
(right scale) shows the space density of AGN with X--ray luminosity 
$log L_X > 44.5\ergs$ as a function of redshift. Above this luminosity 
the XLF is a steep power law for all observed redshift shells and the
data in Fig. \ref{ZLF} have been determined by assuming the slope to be 
independent of redshift and fitting the normalization of the XLF
(Miyaji, Hasinger \& Schmidt, 1998). The X--ray data do not  
show a significant decrease of the space density of high--redshift,
high--luminosity
X--ray selected AGN and appear to be marginally inconsistent with the 
optical and radio determinations (Miyaji, Hasinger \& Schmidt, 1998). 
However, the X--ray surveys still suffer
from small sample sizes at high redshift (see Fig. \ref{RXZ}), so that
significantly larger solid angles have to be covered to a similar depth 
and optical completeness as the ultradeep HRI survey 
in order to get a clear picture of the AGN density at high redshift.
The array of planned Chandra and XMM surveys 
in other fields than the Lockman Hole will be of great help in this respect.

\subsection{AGN contribution to the star formation history}

The star formation history in the universe out to redshifts of
four has been studied in the last few years by optical and
NIR observations using ground--based telescopes and deep photometric
surveys with the Hubble Space Telescope (see e.g. Blain et al., 1998, 
for a recent review). The open circles in Fig. \ref{ZLF} show
the compilation of the most recent observational determinations 
of the optical/UV star forming rate (SFR) by Blain et al. (left scale),
which suggest that star formation peaked at a redshift around 1--2. 
These data points, however, have to be regarded as lower limits of the true 
SFR because much of the light emitted in star bursts can be significantly
obscured. Recently the far--infrared/sub--mm extragalactic background
light (FIB), the equivalent of the X--ray background at very long
wavelengths, has been discovered (Puget et al. 1996; Fixsen et al. 1998).
Deep SCUBA surveys have detected a population of optically faint galaxies, 
luminous in the sub--mm band, which could produce a significant
fraction of the FIB signal (Smail, Ivison \& Blain 1997;
Hughes et al. 1998; Barger et al. 1998). Source counts of dusty 
galaxies and AGN in the sub--mm band are strongly weighted towards high 
redshift because of the large negative K--correction of the very steep dust 
spectra (Blain \& Longair 1993). If all of the FIB should be due to     
star forming processes, a large population of strongly obscured
star bursting galaxies would be missing from the optical/UV surveys at 
high redshifts. The dashed line in figure \ref{ZLF} sketches one of
the SFR models by Blain et al., that is able to produce all of the
FIB by early star formation. These models still have some drawbacks, 
however, because this massive early star formation would likely 
overproduce the heavy elements and consume a large fraction of 
all baryons in the universe into stars (Blain et al. 1998). 

It is interesting to note that the star formation rate required to 
produce the FIB light has a cosmic history which is very similar to the
dependence of the AGN space density on redshift (see Fig. \ref{ZLF}). 
Could it be, that active galactic nuclei contribute significantly
to the faint sub--mm source population? The X--ray background population
synthesis models have recently been used by Almaini et al. (1999) to 
predict the AGN contribution to the sub--mm background
and source counts. Depending on the assumptions about cosmology and
in particular on the AGN space density at high redshifts (see above)
they predict that a substantial fraction of the sub--mm source counts 
at the current SCUBA flux limit could be associated with active galactic 
nuclei. Interestingly, the first optical identifications of SCUBA sources 
indicate a significant AGN contribution (Smail et al. 1999).

Another, largely independent line of arguments leads to the conclusion
that accretion processes may produce an important contribution to the
extragalactic background light. Dynamical studies (e.g. Magorrian et al.
1998) come to the conclusion that massive dark objects, most
likely dormant black holes, are ubiquitous in nearby galaxies. There
is a correlation between the black hole mass and the bulge mass of 
a galaxy: $M_{BH} \approx 6 \times 10^{-3} M_{Bulge}$. Since gravitational 
energy release through standard accretion of matter onto a black hole is
producing radiation about 100 times more efficiently than the thermonuclear
fusion processes in stars, the total amount of light produced by accretion 
in the universe should be of the same order of magnitude as that produced by
stars. A more detailed treatment following this argument comes to the 
conclusion that the AGN contribution should be about 1/5 of the stellar      
light in the universe (Fabian \& Iwasawa, 1999).

Regardless of whether the FIR light of AGN is from dust heated by stellar
processes or by accretion onto the massive black hole, these studies 
indicate that a large contribution to the light emission history in 
the early universe could come from sources associated with AGN,
which are most easily pin-pointed by sensitive X--ray observations.
Future joint sub--mm/X--ray deep surveys will therefore be a very 
powerful tool to disentangle the different processes dominating the universe in
the redshift range $2<z<5$.

\begin{acknowledgements}
The ROSAT project is supported by the Bundesministerium
f\"ur Forschung und Technologie (BMFT), by the National
Aeronautics and Space Administration (NASA), and the Science
and Engineering Research Council (SERC). 
This work has been supported in part by the DLR (former DARA GmbH) 
under grant 50~OR~9403~5 (G.H and I.L.) and 
by NASA grants NAG5--1531 (M.S.), NAG8--794, NAG5--1649,
and NAGW--2508 (R.G.). G.Z. acknowledges partial support by the Italian 
Space Agency (ASI) under contract ARS-96-70.
\end{acknowledgements}

\end{document}